\documentclass[useAMS,usenatbib]{mn2e}
\def\deg{$^\circ$}
\usepackage{subfigure}
\usepackage{graphicx}
\title[Periodic Radio Pulses from J1732-3131]
{Pulsed Radio Emission from the Fermi-LAT Pulsar J1732-3131: Search and A Possible Detection at 34.5 MHz}
\author[Y. Maan, H. A. Aswathappa and A. A. Deshpande]
{Yogesh Maan$^{1,2}$ \thanks{E-mails: yogesh@rri.res.in (YM), 
aswath@rri.res.in (HAA) and desh@rri.res.in (AAD)} 
H. A. Aswathappa$^{1}$ and Avinash A. Deshpande$^{1}$\\
$^{1}$Raman Research Institute, Bangalore 560080, India\\
$^{2}$Joint Astronomy Programme, Indian Institute of Science, Bangalore 560012, India}

\begin{document}
\maketitle
\label{firstpage}
\begin{abstract}

We report our search for and a possible detection of periodic 
radio pulses at 34.5 MHz from the Fermi-LAT pulsar J1732-3131. 
The candidate detection has been possible in only one of the 
many sessions of observations made with the low-frequency array 
at Gauribidanur, India, when the otherwise radio weak pulsar 
may have apparently brightened many folds. The candidate dispersion 
measure along the sight-line, based on the broad periodic profiles 
from $\sim$ 20 minutes of data, is estimated to be $15.44\pm0.32$ pc/cc.
We present the details of our periodic \& single-pulse search, and 
discuss the results and their implications relevant 
to both, the pulsar and the intervening medium.
\end{abstract}
\begin{keywords}
pulsars: general -- pulsars: individual (J1732-3131)
\end{keywords}
\section{Introduction}
The Fermi Large Area Telescope (LAT) with its large effective
area of $\sim 9500\; {\rm cm^2}$, large field of view ($\sim 2$ sr),
narrow point spread function $\sim 0.6^{\circ}$ at $1$ GeV
(normal incidence) and broad energy range (20 MeV - 300 GeV),
is a highly sensitive instrument for $\gamma$-ray observations.
Because of this vast improvement in sensitivity compared to
past $\gamma$-ray missions, an extraordinary increase has been
seen in the number of pulsars detected in gamma rays
Apart from the detection of gamma-rays from a large number
of known pulsars, it was for the first time 
possible to perform blind searches for pulsars in $\gamma$-rays. 
These blind searches have so far discovered 33 
$\gamma$-ray pulsars from the data recorded by the Fermi-LAT
\citep[][]{Abdo09,Saz10,Pletsch12}.
There are mainly two kinds of competing theoretical models
for the high energy emission from pulsars, viz. the Polar
cap models and the Outer gap models. The ratio of
radio-quiet to radio-loud $\gamma$-ray pulsars is an important
discriminator between these two types of models. The LAT-pulsars 
discovered in blind searches need not necessarily be radio-quiet, 
and hence need to be searched for radio pulsations.
Despite deep radio follow-up searches, 
so far only four of these LAT-discovered pulsars could be detected 
\citep[J1741-2054, J2032+4127, J1907+0602 \& J0106+4855;][]
{Camilo09,Abdo10,Pletsch12}, suggesting that the radio-quiet pulsar 
population might actually be quite large. However, all the deep 
follow-up searches were carried out at high radio frequencies 
\citep[around 1 GHz and above, except a few at 300 MHz ;][]
{Ray11,Pletsch12},  and the lower frequency domain still remains 
relatively unexplored.
\par 
At low radio frequencies the emission beams become wider 
(suggesting the so called radius-to-frequency mapping (RFM) 
of pulsar radio beams) increasing the probability of our
line-of-sight passing through the emission beam. A classic
example is B0943+10, which is detected only below $\sim 1$ GHz
because our line-of-sight misses the relatively narrow emission
beam at higher frequencies almost completely \citep[see, for example,][]{Weisberg99,DR01}. Also, recent study \citep{Ravi10}
suggests that for high-$\dot E$ (young) and millisecond pulsars,
radio emission beam widths are comparable to those of $\gamma$-ray beams. Therefore
such pulsars, if already detected in gamma-rays, can be expected to 
be beaming towards us. Given the widening of radio 
beams at low frequencies even for 
normal pulsars, follow-up searches 
of these $\gamma$-ray pulsars at very low 
frequencies ($< 100$ MHz) could also be revealing. Any detection 
would open an otherwise rare possibility of studying these 
objects at such low radio frequencies. 
On the other hand, a non-detection might provide 
stringent upper limits on the radio flux from these sources.
Thus search for radio emission from LAT-discovered pulsars 
at low frequencies is essential for completeness before 
discriminating between the polar cap and outer gap models.
\par
Many of the pulsars discovered by the Fermi-LAT lie in sky 
regions surveyed using Gauribidanur Telescope at 
34.5 MHz in the years 2002 to 2006. We have searched the 
archival data of this survey, for periodically pulsed or 
any transient signal along the direction of 17 of these 
LAT-PSRs. Most of these remained undetected in both kind 
of searches. A second round of search and analysis using 
the same data but with relaxed thresholds is in progress. 
We present here detection of periodic radio pulses 
along the direction of the LAT-PSR J1732-3131 \citep[][]{Abdo09},
corresponding to period of $0.19652\pm0.00003$ seconds and dispersion
measure of $15.44\pm0.32$ pc/cc. The results along the direction 
of other LAT-discovered pulsars will be reported elsewhere.
\par
Section 2 describes the observations, and presents the result
of the search carried out in the direction of pulsar J1732-3131. 
In section 3, we discuss the detection significance and present the 
estimates of flux density \& distance to the 
pulsar, followed by conclusions in section 4.
\section[]{Search for Single Bright and Periodic Pulses}
\subsection[]{Observations}
Observations used here were made as a part of pulsar/transient
survey, using the East-West arm of the radio telescope at 
Gauribidanur, India. The beam-widths (full width at half-power) of 
this arm are 21' and 25\deg $sec(zenith-angle)$ in right ascension 
and declination respectively. The effective 
collecting area offered by this arm is about 12000 m$^2$ at 
the instrumental zenith ($+14^{\circ}.1$ declination). 
In each observing session, raw signal voltage sequence was directly 
recorded at the Nyquist rate (with 2-bit, 4-level quantization) using 
Portable Pulsar 
Receiver\footnote{http://www.rri.res.in/$\sim$dsp$\_$ral/ppr/ppr$\_$main.html (Deshpande, Ramkumar, Chandrasekaran \& Vinutha, in preparation)}(PPR)
for about 20 minutes, while tracking the source.
For further details about the telescope we refer the 
reader to Deshpande, Shevgaonkar and Sastry (1989).
\par
In the direction of the pulsar J1732-3131, data are available 
from 10 observing sessions of the above survey. The voltage 
time-sequence from each of these observing sessions was, in 
the off-line processing, Fourier transformed in blocks of 
512 samples, resulting in dynamic spectrum with
256 spectral channels across 1.05 MHz bandwidth centred around 34.5 MHz, and 
temporal resolution of $\sim1.95\; ms$ (after averaging 
8 successive raw power-spectra).
These resultant data in filter-bank format were subjected to 
two types of searches described in the following sub-sections.
\subsection{Search for Single Bright Pulses}
Periodic signals from pulsars at such low frequencies are 
generally very weak, and detectable only from a small fraction 
of the total pulsar population. However, it is not uncommon to find 
the strength of a single pulse, or a part of the pulse, increase 
many folds from its average (e.g. in the form of explicit excursion 
as in giant pulses/radiation spikes, or as a part of general
sub-pulse level intensity fluctuations). How well such signals
can be differentiated from mere noise fluctuations, or those due 
to radio frequency interference (RFI), depends not only on their 
relative strength, but also
on their arrival times showing any specific correspondence
relative to the expected pulse arrival, and if manifestation of 
their propagation through the intervening medium, as for distant 
astronomical sources, is clearly evident.
\par
The time of arrival of a pulsed signal 
varies across the frequency band because of the interstellar
dispersion, a characteristic that serves as a primary identifier
for signals from distant astronomical sources. The relative delay 
at a frequency $\nu$ with respect to a reference frequency $\nu_0$ 
is given by: $\Delta t \rm{\;(ms)} = 
4.15\times10^6\times DM\times(\nu^{-2}-\nu_0^{-2})$; where $\nu$ and 
$\nu_0$ are in MHz, and DM (Dispersion Measure in units of pc/cc) is 
the column density of electrons between the observer and the source, 
along the line of sight.
\par

\begin{figure}
  \begin{center}
    \includegraphics[scale=0.35,angle=-90]{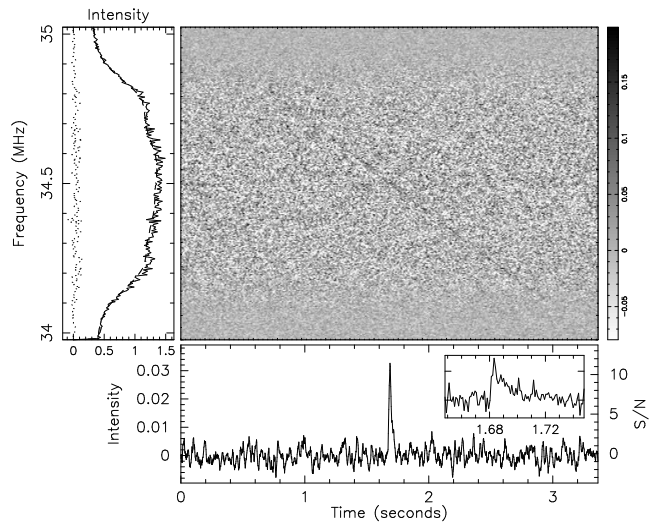}
    \caption{{Dynamic spectrum of the 10 single pulses aligned and 
averaged together (to enhance S/N) is shown in the main panel. 
The data are smoothed by a rectangular window of width about 
16 ms. The bottom sub-plot shows the pulse obtained after correcting 
for the delays corresponding to DM = 15.55 pc/cc. The left panel shows 
the average spectrum (dashed), the spectrum corresponding to the 
pulse-peak (solid) and the residuals (dotted) on the same scale.
The inset in the bottom subplot shows average dedispersed 
pulse without any smoothing.}
}
  \label{spulses_DynSpec}
  \end{center}
 \end{figure}

Search for single bright pulses looks for this dispersive signature
across the observing band at each of the trial DMs, and reports its 
significance, in case of detection above a given threshold 
\citep[see for example,][]{CM03}. 
Further considerations are necessary to take in to account the effect of 
scattering in the intervening medium. Even for moderate DMs, 
the apparent pulse shape at such low frequencies is expected to 
be dominantly affected by the interstellar scattering. 
The resultant effect is usually modeled as a convolution of 
intrinsic pulse profile with a truncated (one-sided) exponential function. 
Thus, to enable optimum detection, a given dedispersed time-sequence
was smoothed (match-filtered) with a truncated exponential 
template, before subjecting to detection criterion. 
This effectively provided a search also in the apparent pulse-width,
when the templates of different widths were used in systematic trials.
In our search for single pulses, with different trial DMs and smoothing 
widths, we found the highest number of pulses (10) exceeding our detection 
threshold of $4.5\,\sigma$ at a trial DM of 15.55 pc/cc, and a smoothing 
width of 8 samples, in contrast with the expectation of less than one 
pulse from noise excursions (specifically, 0.27 for the sequence of 
77950 independent samples).
These single pulses vary in their strengths from $4.5\,\sigma$
to $5.2\,\sigma$, and were found to be distributed more or less uniformly 
throughout the total duration of the observation. When these 10 individual 
pulses (belonging to the DM-bin of width $0.29$ pc/cc) were aligned and 
averaged together, the signal-to-noise ratio (S/N) improved as expected, 
providing a refined estimate of DM ($15.55\pm 0.04$ pc/cc). 
Figure~\ref{spulses_DynSpec} shows this average, where the dispersive nature 
of the resultant pulse in the dynamic spectrum is apparent.
But, we need to remind ourselves that 
however appealing the appearance of such an average dispersed pulse may be, 
it is to be viewed with caution. A relevant illustration, though in a 
different context, can be found in \citet{Gold08}, where constructive 
alignments and co-additions are shown to result in spurious peaks which
are entirely due to random noise. The present situation is potentially 
no different from the above example, except that now we might have
noise manifestations that are pre-selected on the criteria of better match 
with certain dispersive characteristics. When many such realizations are 
combined, as in our average of 10 pulses, naturally the result will
show the same dispersive pattern even more clearly.
The only parameter which will distinguish 
between possible spurious pattern and the real signal, if any, is the 
S/N of the resultant pattern. Hence, to assess the real significance 
of the resultant average pulse, the above procedure was repeated a number
of times, in each using 10 strongest pulses at an arbitrarily chosen DM 
(understandably, this required us to reduce the threshold). 
The corresponding average dedispersed pulse was found to be consistently 
weaker than that in figure~\ref{spulses_DynSpec}. However, the difference 
was not very significant (S/N lower only by 1 or 2). This is not very 
surprising considering the fact that S/N of 4 out of the 10 pulses (at the 
candidate DM of 15.55 pc/cc) are only marginally above our detection threshold
(a higher detection 
threshold of $5 \, \sigma$ would have left with us only 2 excursions which 
are too few to seek a refined DM). Given this poor 
distinction between the average of ten pulses at the 
candidate DM and those at other DMs, the confidence with which 
inferences can be drawn about their possible association with the pulsar 
(assessed through their apparent longitudes), their average width 
and a refined DM they may imply, is rather limited, and these results
are to be viewed with due caution\footnote{Our detection 
threshold is admittedly and deliberately kept low ($4.5 \,\sigma$) compared 
to commonly used thresholds, to minimize the probability of `miss', which 
of course increases the false-alarm rate. Hence, there is a finite probability 
that some or most of these excursions may not correspond to the signal we are 
looking for. We note, however, that these observations were conducted at late 
night hours, when possible RFI is at its minimum, if not absent. 
The fact that the excursions span such a narrow range above our particularly
low detection threshold, suggests to us absence of contamination from RFI,
narrowing the causes to random noise or real signal. Quite independently, 
our pre-search processing looks for and excises any significant RFI 
that is narrow in bandwidth and/or in time. And, our procedure of normalizing
the total power at each sample of the dynamic spectrum, removes all signals
that are broadband {\itshape and} are of non-dispersive nature. 
Only those RFI which
carry a dispersive signature, necessarily broadband, can escape this scrutiny.
The ``swept-frequency'' RFI can indeed sometimes mimic the dispersive 
signature, normally identified with astronomical origin. However, 
the times of arrival of these ``swept-frequency'' RFI are generally 
linearly proportional to the frequency (as against the inverse-square 
dependence in case of astronomically originating signals). As part 
of further critical assessment, the data were dedispersed with linear 
delay gradients ($\Delta t \propto \nu$) spanning similar delay ranges, 
and the results were compared with those using $\nu^{-2}$ law. 
As would be expected for astronomical signal, the significance of 
detection is indeed lower for the linear chirp.}.
%
%
\subsection{Search for Periodic Pulsed Signal}
\begin{figure}
  \begin{center}
    \includegraphics[scale=0.5,angle=-90]{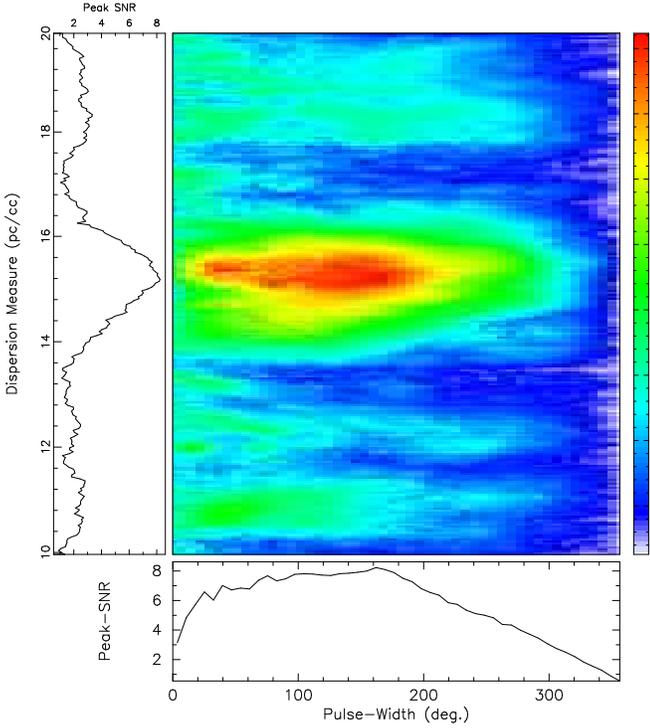}
    \caption{The body of the figure shows the peak-S/N in the 
folded profiles corresponding to each combination of the trial values 
of smoothing-width and DM. The plateau defining the preferred combinations 
is apparent in the range $\sim80$-$180$ degrees in smoothing-width, 
and $\sim14.8$-$15.6$ pc/cc in DM. For ease of viewing the typical variations, 
the left and bottom panels provide  vertical and horizontal 
cuts at smoothing-width=$160^{\circ}$ and DM=$15.19$ pc/cc
through the data plotted in the central panel.}
  \label{j1732_detection}
  \end{center}
 \end{figure}

Since the rotation ephemeris for the LAT-PSR J1732-3131 is 
known \citep{Ray11}, folding the 20 minutes time sequence over 
the rotation period (after corrections for the barycentric 
motion of the Earth) was performed for each of the 
frequency channels to enhance S/N. Then, this ``folded'' 
dynamic spectrum was used to search for any significant 
dispersed pulse-profile. Peak S/N above a threshold was used 
as the criteria for detecting the candidates. However, highest 
S/N is expected when effective time resolution becomes equal to 
the pulse-width, so that all the flux in the pulse is integrated 
in just one sample. This becomes particularly important for a 
candidate having a large pulse-width (a number of reasons 
contribute to the apparent width; e.g., prominent effect of 
interstellar scattering at low frequencies, intrinsic pulse-width 
etc.). An optimum search across trial pulse-widths was thus 
carried out in each of the dedispersed-folded profiles. For
completeness, we also extend the search in the period domain 
(although over a narrow range of period offsets). In 
figure~\ref{j1732_detection}, we present a two-dimensional cut
(DM and pulse-width) of the results of 3-D search procedure
mentioned above, when applied to data from one of the observing 
sessions in 2002. 

\begin{figure}
  \begin{center}
    \includegraphics[scale=0.5,angle=-90]{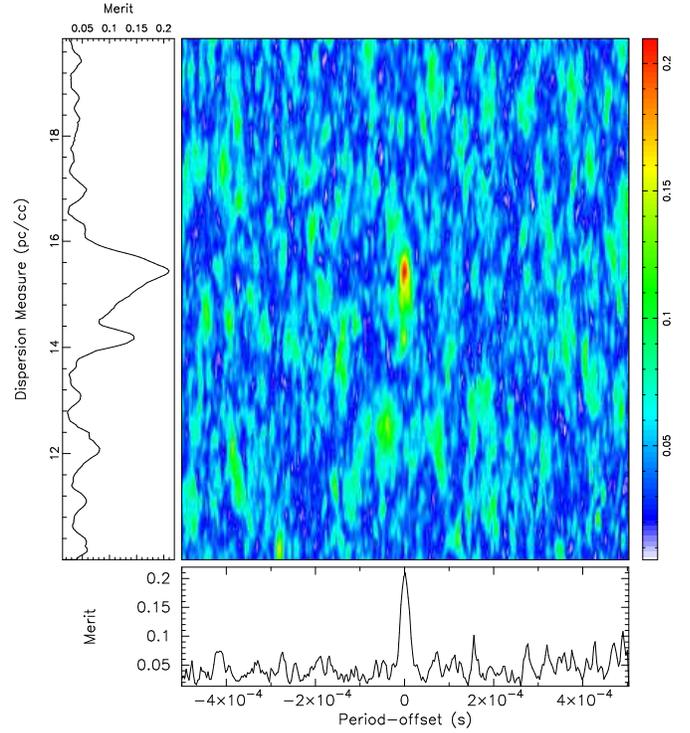}
    \caption{{The main panel shows the significance of the folded 
profile as a function of DM and period. A narrow range of period-values 
around the actual period has been chosen. For the best-fit period, the 
figure-of-merit variations as a function of DM are shown in the left 
subplot. Corresponding to the best-fit DM ($15.44$ pc/cc), the bottom 
subplot shows the variations of figure-of-merit as a function of the 
period-offset.}}
  \label{folded_P0_DM}
  \end{center}
 \end{figure}

Clearly, a broad pulse is detected at DM of $15.2\pm0.4$ pc/cc;
see the S/N profile in the bottom panel, where a broad S/N peak is 
seen in the range $\sim80$-$180$ degrees, indicating structures
within the pulse on these scales.
We note however that the S/N of the peak in the average profile
as a figure of merit can lead to large errors in estimating the 
true DM when the individual frequency channel profiles have poor
S/N, as is the case presently. There is also a related bias towards
compactness of the average profile while searching for best-fit DM.
We therefore use sum-of-squares (SSQ) of average intensities across
the profile as a figure of merit. The significance of the average
profile assessed in this manner and as a function of DM and 
period-offset is shown in figure~\ref{folded_P0_DM}. The isolated
peak in figure of merit is striking and allows a less biased 
estimate of DM ($15.44\pm0.32$ pc/cc), which we adopt in further
discussion. 
Also, the period of 0.19652(3) seconds, suggested
by this isolated peak, is consistent with that extrapolated
from the available ephemeris \citep[][]{Ray11}, to the epoch of our 
observation (0.19652125(2) seconds).
\par
 
\begin{figure}
  \begin{center}
    \includegraphics[scale=0.35,angle=-90]{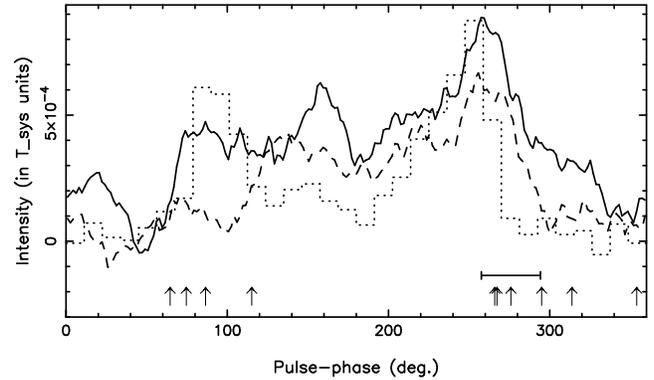}
    \caption{{Average pulse-profiles at 34.5 MHz (solid \& dashed; 
corresponding to DM = 15.44 pc/cc and 15.19 pc/cc respectively),
a scaled version of the gamma-ray pulse-profile (dotted) and the positions 
of peaks of 10 bright pulses (arrow-marks) are shown together for ready
comparison. The radio profiles were smoothed by a 25 degree wide 
window, and manually aligned with the gamma-ray profile. The horizontal 
bar denotes the average width of the bright pulses, which, on average, 
are about 200 times brighter than the peak intensity in the average profile.}}
  \label{j1732_profile}
  \end{center}
 \end{figure}

In figure~\ref{j1732_profile}, the average profiles corresponding to 
the above mentioned two figures-of-merit are shown (peak S/N based : 
dashed-line; SSQ based : solid-line) for ready comparison. Both the 
profiles have been smoothed by only 25 degrees wide window, to retain 
the primary details in the profile, although the S/N would be less than
optimum. We would like to 
emphasize here that although folding of random noise can also produce 
impressive profiles \citep[as exemplified in ][]{Ramach98}, peak amplitude 
in such profiles would still be constrained within the noise limits (i.e. 
full swing within $3$ to $4\,\sigma$ of the noise). In the present case, 
the peak-S/N in the average profile (figure~\ref{j1732_profile}; solid-line) 
is estimated to be 7. Here, the used value of noise standard deviation (in 
units of T$_{\rm{sys}}$) is deduced directly from the time-bandwidth product, 
assuming only $75\%$ of the total bandwidth. If we were to use a more 
optimistic estimate, using the rms-deviation computed from the dedispersed 
time sequence, the peak-S/N would exceed 9. It is clear that even the 
conservative estimate of the peak-S/N (7) is well outside the above mentioned 
threshold, and hence the claimed signal in the average profile can not be 
dismissed as mere manifestation of random noise.

\subsection{Confirmatory Checks}
To further assess the credibility of the above detection, the
folded profiles prepared using the two halves of the frequency 
band separately, and those using alternate frequency channels 
were examined for the dispersion signature correspondingly.
In another sanity check, a two-dimensional search in DM \& 
pulse-width was carried out on the folded profiles made using 
only odd and even period-numbers separately. Both independently 
showed peak around DM$\sim$15.4 pc/cc. We also carried out these 
searches on two fields observed on the same day just 20 minutes (in 
right ascension) prior to and after the field containing the candidate. 
No sign of dispersed signal around this period was found in either of 
the fields. This makes it unlikely that the above detection of a periodic 
signal could have been a result of some man-made or system originated signal.
\par
Intriguingly, for rest of the 9 observing sessions, neither of the two 
kinds of searches showed any significant detection. We will discuss various 
possibilities which may have caused these non-detections, along with the 
implications of parameters determined from our data (table 1) in the 
following section. 
\begin{table*}
 \centering
  \caption{Measured and Derived parameters for PSR J1732-3131}
  \begin{tabular}{lrrlrr}
  \hline
  \hline
  \multicolumn{3}{c}{Known Parameters \citep[from][]{Ray11}} & \multicolumn{3}{c}{Measured and 
  Derived Parameters}\\
 \hline
  Right Ascension, R.A.(J2000.0)	&	&	17:32:33.54	& Dispersion Measure, DM(pc/cc)        		&       &       $15.44\pm0.32$\\
  Declination, Dec. (J2000.0)           &       &       -31:31:23.0     & Best-fit Period, $P_0$(s)                     &       &$0.19652\pm0.00003$\\
  Pulse Frequency, $\nu$ ($s^{-1}$)     &       &       5.08794112      &  Epoch of period (MJD)                        &       &       $52384.9209$\\
  Frequency first derivative, $\dot{\nu}$ ($s^{-2}$)&& -7.2609$\times10^{-13}$& Distance, $d_{PSR}$ (pc)               &       &       $600\pm150$\\
  Epoch of frequency (MJD)              &       &       54933.00        &   Pulse Width, $\Delta\phi$ (degrees)      &       &       $<\,200$\\
\hline
\end{tabular}
\end{table*}
\section[]{Results and Discussion}
In figure~\ref{j1732_profile}, the average profile of the 
LAT PSR J1732-3131 at 34.5 MHz (solid-line) is compared with the 
pulse-profile as seen in gamma-rays (dotted-line) 
\citep[][see their figure 25]{Ray11}. 
The radio profiles have been aligned with $\gamma-$ray profile 
manually, since the accuracy in the estimated DM value
is not adequate enough for absolute phase-alignment. The apparent similarity 
between the profiles at these two extreme ends of the spectrum is striking,
although there are possible differences. At radio frequencies, 
the significant pulsed emission is confined to about $70\%$ of 
the period, with possibly bridging emission between the two gamma-ray 
components (i.e. between 120 to 220 degrees).
\subsection*{Origin of Single Pulses ?}

The distribution of single bright pulses (figure~\ref{j1732_profile};
see the ``arrow-marks'') in
longitude is bimodal rather than uniform, visiting regions near
the leading and trailing components of the main broad pulse in 
the radio-profile. Given that 
1) no obvious association of these single pulses with any of the 
known pulsars is apparent (based on DM),
2) the dispersion measure suggested by these bright pulses
(DM=$15.55\pm0.04$ pc/cc) falls within the error limits of that 
associated with the periodic signal ($15.44\pm0.32$ pc/cc), 
3) the profile shapes at these two DM values are indistinguishable 
(not shown in figure~\ref{j1732_profile}, but assessed separately), and 
4) the position of these pulses are correlated with the outer 
regions of the pulse-window; 
it is difficult to rule out the possibility that these single pulses
share their origin with the periodic signal. 
At the same time, the statistics of 10 pulses is too poor to rule out
chance segregation in longitude. Giant pulses/radiation spikes 
are usually seen to confine themselves within the average pulse window
\citep[][]{L95,Ables97}.
However, in the case of J0218+4232 \citep{Knight06}, the giant pulses
concentrate just outside the rising and trailing edges of the broad
radio pulse, and those regions match the locations of the peaks in 
the high energy (x-ray) profile. If our single pulses are real, then 
their apparent
distribution is reminiscent of the situation in J0218+4232, and the 
peaks in the gamma-ray profile
could correspond to the single pulse locations if the gamma-ray 
profile was shifted by about $180^{\circ}$ compared to what is shown
in figure~\ref{j1732_profile}.
Given this, and the apparent relative brightness of our single pulses, 
the possibility of their being giant pulses can not be ruled out.

\subsection*{Flux Density Estimate}
The half-power beam-width at this declination is almost 
35$^{\circ}$ (in declination). Therefore, we take the sky temperature 
estimates at various points across the beam from the higher 
resolution synthesis sky-map at 34.5 MHz \citep{DU90}, and a
weighted average of these using a theoretical beam-gain pattern
provides us an estimate for the system temperature. For the present
case, it is estimated to be $57000^{\circ}$K (receiver temperature 
contribution is negligible). Using this calibration, and assuming
an effective collecting area of $8700 \; {\rm m^2}$ (in the direction
of the pulsar) we estimate the average flux density (pulse-energy/period)
of this pulsar at 34.5 MHz to be about 4 Jy. 
\subsection*{Non-detection at Other Frequencies and in Other Observing Sessions}
The mean flux density at 34.5 MHz, combined with the upper limit on the 
flux density at 1.4 GHz \citep[0.2 mJy;][]{Camilo09}, suggests a spectral 
index $\alpha\leq -2.7$ assuming no turn-over in the spectrum (a
tighter upper limit given by \citealt{Ray11} from a deeper search 
suggests further steeper spectral index). 

Although such a steep spectrum 
could explain the non-detection of the pulsar at higher frequencies, 
its detection in only one out of ten observing sessions at 34.5 MHz 
necessitates consideration of following possibilities :\\
1. {\bf Extrinsic to the source: }This pulsar may actually be 
emitting below our detection limit, and favorable refractive scintillation 
conditions possibly raised the flux above our detection 
limit during one of our observing sessions.\\ 
2. {\bf Intrinsic to the source: }The source may be an intermittent 
pulsar or a radio-faint pulsar which comes in ``radio-bright'' mode 
once in a while.
\par
In either of the above possibilities, the spectral steepness mentioned 
above would be an overestimate. 
\subsection*{Distance Estimates}
For DM=15.4 pc/cc and location of the pulsar 
(RA=17:32:33.54, Dec=-31:31:23.00), the \citet{CL02} electron density
model (C\&L model) yields\footnote{The typical uncertainty in distance
estimated using the C\&L model, as quoted above, is believed to be 25\%,
however in some cases the estimates can be uncertain even by a factor of 2.}
a pulsar distance ($d_{PSR}$) of $600\pm150$ pc. 
This agrees well with the $d_{PSR}$ estimates of $0.77^{+0.41}_{-0.35}$ kpc 
and $0.86^{+0.49}_{-0.30}$ kpc by \citet{Wang11}, using the correlation 
between the gamma-ray emission efficiency and a few pulsar parameters 
(generation-order parameter and $B_{LC}$).
Alternatively, these distance estimates combined with the model 
electron density along the pulsar direction, give consistent 
DM value. However, the pulse broadening because of the interstellar
scattering as predicted from C\&L model ($87^{+50}_{-35}$ ms),
appears to be over-estimated by a large factor, 
given the narrow widths of bright pulses (about 20 ms or narrower) 
and the width of pulse components in the average pulse-profile.
\par
Follow-up observations at higher frequencies (100 to 300 MHz) and 
with better sensitivity would be useful in improving DM estimate, 
and for confirming the most likely association of the bright pulses 
with the pulsar. If confirmed, the comparison between the pulse profiles 
in the two extreme parts of the electromagnetic spectrum, and the 
location of bright single pulses would provide further insight into 
emission in the form of giant pulses \citep[e.g.][]{Hankins03,Knight06} 
or radiation spikes \citep[e.g.][]{Ables97}.
\section[]{Conclusions}
In this paper, we have presented results of our two kinds of searches
(at 34.5 MHz) for 
radio pulses in the direction of the LAT pulsar J1732-3131, and report in 
detail an intriguing detection of periodic pulsed signal 
at DM of about $15.44$ pc/cc. The possible reasons for its non-detection, 
in our other observing sessions and at higher frequencies, and 
related implications are discussed. 
The DM based distance estimate, using Cordes \& Lazio electron 
density model, matches well with earlier estimates based on $\gamma$-ray
emission efficiency. These results also demonstrate the potential and 
the importance of such searches/surveys at low radio frequencies.
We hope our estimate of DM along with the other results
will help follow-ups at suitable higher frequencies.
\section*{Acknowledgments}
We thank our referee, Fernando Camilo, for his comments 
and suggestions which have helped in improving our manuscript.
The Gauribidanur radio telescope is jointly operated by 
the Raman Research Institute and the Indian Institute of 
Astrophysics. We gratefully acknowledge the support from 
the observatory staff. YM is thankful to 
Harsha Raichur, Nishant Singh and Wasim Raja for useful 
discussions and comments on the manuscript.

\label{lastpage}
\end{document}